\documentclass[aps,floats]{revtex4}
\usepackage{amsmath,amssymb}
\usepackage{graphicx,epsfig}
\usepackage[greek,english]{babel}
\usepackage{bbold}
\begin{document}
\bibliographystyle {plain}

\pdfoutput=1
\def\oppropto{\mathop{\propto}} 
\def\opsimeq{\mathop{\simeq}}
\def\opoverderline{\mathop{\overline}}
\def\operarrow{\mathop{\longrightarrow}}
\def\opsim{\mathop{\sim}}

\def\fig#1#2{\includegraphics[height=#1]{#2}}
\def\figx#1#2{\includegraphics[width=#1]{#2}}

%\newcommand{\fig}[2]{\epsfxsize=#1\epsfbox{#2}} \reversemarginpar 

%%%%%%%%%%%%%%%%%%%%%%%%%%%%%%%%%%%%%%%%%%%%%%%%%%%%%%%%%%%%%%%%%%%%%%%%%%%%
\title{ Topological phase transitions in random Kitaev $\alpha$-chains  } 

%%%%%%%%%%%%%%%%%%%%%%%%%%%%%%%%%%%%%%%%%%%%%%%%%%%%%%%%%%%%%%%%%%%%%%%%%%%%

\author{ C\'ecile Monthus }
 \affiliation{Institut de Physique Th\'{e}orique, 
Universit\'e Paris Saclay, CNRS, CEA,
91191 Gif-sur-Yvette, France}

\begin{abstract}
The topological phases of random Kitaev $\alpha$-chains are labelled by the number of localized edge Majorana Zero Modes. The critical lines between these phases thus correspond to delocalization transitions for these localized edge Majorana Zero Modes. For the random Kitaev chain with next-nearest couplings, where there are three possible topological phases $n=0,1,2$, the two Lyapunov exponents of Majorana Zero Modes are computed for a specific solvable case of Cauchy disorder, in order to analyze how the phase diagram evolves as a function of the disorder strength. In particular, the direct phase transition between the phases $n=0$ and $n=2$ is possible only in the absence of disorder, while the presence of disorder always induces an intermediate phase $n=1$, as found previously via numerics for other distributions of disorder.

\end{abstract}

\maketitle

\section{ Introduction}

Many one-dimensional quantum models involving $N$ quantum spins $S=1/2$ (see the Appendix) or $N$ spinless Dirac fermions can be reformulated in terms of $2N$ of Majorana operators $\gamma_j$ 
that are hermitian $\gamma_j^{\dagger}=\gamma_j$, square to the Identity $\gamma_j^2=  \mathbb{1}$ and anti-commute with each other. For models respecting the Time-Reversal-Symmetry $T$ defined by its action \cite{kitaevchain,kitaevfid} 
\begin{eqnarray}
T i T^{-1} && =  -i
\nonumber \\
T \gamma_{2j-1} T^{-1} &&=  \gamma_{2j-1}
\nonumber \\
T \gamma_{2j} T^{-1} && = - \gamma_{2j}
\label{time}
\end{eqnarray}
it is actually useful to relabel the Majorana operators with the flavors $a$ and $b$ to stress their different behaviors with respect to $T$
\begin{eqnarray}
 \gamma_{2j-1} && = a_j
\nonumber \\
 \gamma_{2j} && = b_j
\label{timeab}
\end{eqnarray}
Among the Hamiltonians respecting also the the total parity 
\begin{eqnarray}
P   = i^N \gamma_1 \gamma_2 \gamma_{3} \gamma_{4}   ... \gamma_{2N-1} \gamma_{2N}
\label{paritytotal}
\end{eqnarray}
the simplest ones are the free-fermionic Kitaev $\alpha$-chains \cite{kitaevchain,kitaevfid,pollmann_unified,pollmann_topology,verresen}
\begin{eqnarray}
 H_{\alpha}= i \sum_m b_m K_{m,m+\alpha}  a_{m+\alpha}
\label{halpha}
\end{eqnarray}
where the diagonalization simply corresponds to the pairing of $(b_n,a_{m+\alpha})$, even for random couplings $K_{m,m+\alpha}$.
Then the possible edge Majorana zero modes are localized on single sites and are thus obvious.
 In particular, $n=\vert \alpha \vert$ counts the number of Majorana Zero Modes of type $a$ or $b$
located near the two edges. For instance, 
$\alpha=0$ corresponds to $n=0$ zero modes, 
$\alpha=1$ corresponds to the single zero mode $a_1$ on the left and the single zero mode $b_N$ on the right,
$\alpha=2$ corresponds to two zero modes $(a_1,a_2)$ on the left and two zero modes $(b_{N-1},b_N)$ on the right, etc.

Many interesting random free Majorana models respecting the $(P,T)$ symmetries,
corresponding either to standard quantum spin models (see the Appendix) 
or to spinless Dirac fermions superconducting models \cite{kitaevchain,kitaevfid,motrunich},
 can be rewritten as a linear combinations with a finite number of values of $\alpha$ \cite{pollmann_unified,pollmann_topology,verresen}
\begin{eqnarray}
H=\sum_{\alpha}  H_{\alpha} =  i \sum_m b_m \left( \sum_{\alpha} K_{m,m+\alpha}  a_{m+\alpha} \right)
\label{hbafree}
\end{eqnarray}
 The topological phases are characterized by the number of edge localized Majorana Zero Modes, 
while the phase transitions between them correspond to delocalization transitions for these edge 
Majorana Zero Modes.
For instance, the linear combination of the three values $\alpha=0,1,2$ 
\begin{eqnarray}
H= H_{0} +H_1+H_2 =  i \sum_n b_n \left( \sum_{\alpha=0}^2 K_{n,n+\alpha}  a_{n+\alpha} \right)
\label{h012}
\end{eqnarray}
has been much considered recently, mostly with homogeneous couplings \cite{pollmann_unified,pollmann_topology,niu2012,ardonne2018,lieu2018}
but also with random couplings  \cite{lieu2018},
in order to analyze the phase diagram of the three possible phases $n=0,1,2$
and the phase transitions between them.
The standard method to analyze the phase diagram of the model of Eq \ref{h012} \cite{pollmann_unified,pollmann_topology,niu2012,ardonne2018,lieu2018},
or more generally of the free-Majorana-models of Eq. \ref{hbafree},
is the transfer matrix computation of the Majorana Zero Modes to characterize their localization properties near the edges
\cite{motrunich,karevski2000,degottardi2011,degottardi2012,degottardi2013,gergs2016,hedge2016,kawabata2017,chitov2018,habibi2018,habibi2018bis}.
In the random case, the analysis of the Lyapunov exponents of the product of random transfer matrices
\cite{motrunich,karevski2000,degottardi2012,degottardi2013,gergs2016,hedge2016,kawabata2017,lieu2018}
is thus based on the methods that have been developed in the field of Anderson localization
and other one-dimensional disordered models (see the books \cite{luck,vulpiani} and the 
more recent Lectures Notes \cite{tourigny} as well as references therein).

The goal of the present paper is to compute the exact phase diagram between the topological phases $n=0,1,2$
for the Hamiltonian of Eq. \ref{h012} for a special case of disorder.
The paper is organized as follows.
In section \ref{sec_zero}, we recall the general method to compute zero modes in random Kitaev $\alpha$-chains,
and the explicit application for arbitrary disorder to the three cases $(H_0+H_1)$, $(H_0+H_2)$ and $(H_1+H_2)$ 
containing only two values of $\alpha$ among the three values of Eq. \ref{h012}.
In section \ref{sec_h3}, we focus on the topological phases of the random Hamiltonian
$ H_{0} +H_1+H_2$ of Eq. \ref{h012},
where the localizations properties of the 
Majorana zero modes can be then obtained via the product of $2 \times 2$ random transfer matrix and via the Riccati recurrence method.
The explicit solution for a special type of Cauchy disorder is given in section \ref{sec_cauchy},
in order to analyze the changes of the phase diagram as a function of the disorder strength.
Our conclusions are summarized in \ref{sec_conclusion}.
Appendix \ref{sec_app} contains a short dictionary between Majorana and quantum spin chains.

\section{ Reminder on edge Zero Modes in random Kitaev $\alpha$-chains }

\label{sec_zero}

\subsection{ General method to compute edge Majorana Zero Mode} 

For the general $PT$-symmetric quadratic models of Eq. \ref{hbafree},
where each Majorana fermion of a given flavor, respectively $a$ or $b$,
interacts only with Majorana fermions of the other flavor, respectively $b$ or $a$,
the zero modes can be also separated into the two flavors,
and can be constructed from some linear combination of the Majorana fermions of flavor $a$ only
\begin{eqnarray}
A =  \sum_j u_j a_j
\label{azero}
\end{eqnarray}
or from some linear combination of the Majorana fermions of flavor $b$ only
\begin{eqnarray}
B =  \sum_j v_j b_j
\label{bzero}
\end{eqnarray}

The linear combination $A$ of Eq. \ref{azero}
will be a Majorana fermion if the coefficients $u_j$ are real 
 $u_j^*=u_j$ and if they satisfy 
the normalization condition
\begin{eqnarray}
 \sum_j u_j^2 =1
\label{normazero}
\end{eqnarray}
This Majorana fermion $A$ will be a Zero Mode
if the commutator with the Hamiltonian vanish
\begin{eqnarray}
0=[H,A]=  2 i  \sum_j u_j \sum_{\alpha} b_{j-\alpha} K_{j-\alpha,n}  
= 2i \sum_{m} b_m \left(   \sum_{\alpha} K_{m,m+\alpha}  u_{m+\alpha}\right)
\label{eqzeroa}
\end{eqnarray}
This condition yields the following recursion for the coefficients $u_j$ that should be satisfied for any $m$
\begin{eqnarray}
0 =  \sum_{\alpha} K_{m,m+\alpha}  u_{m+\alpha}
\label{recun}
\end{eqnarray}
i.e. the coefficients $u_j$ correspond to a right eigenvector of the coupling matrix $K_{m,m+\alpha}  $ associated to the zero eigenvalue.

Of course, on can apply the same analysis for the Majorana Zero Mode $B$ of flavor $b$ of Eq. \ref{bzero},
 and one obtains that the real normalized coefficients $v_j$ correspond to a left eigenvector of the coupling matrix $K_{m,m+\alpha}  $ associated to the zero eigenvalue,
but this will not be discussed further here in order to avoid repetitions.

\subsection{ Example $H=H_0+H_1$  }

For the case
\begin{eqnarray}
H=H_0+H_1= i \sum_{n=1}^N b_n K_{n,n}  a_{n} +  i \sum_{n=1}^{N-1} b_n K_{n,n+1}  a_{n+1}
\label{halpha01}
\end{eqnarray}
corresponding to the random quantum spin Ising chain (see Eqs \ref{halpha0} and \ref{halpha1} in Appendix \ref{sec_app}),
the competition is between the phase $n=0$ with no zero mode (when $H_0$ dominates over $H_1$)
and the phase $n=1$ with one zero mode of flavor $a$ localized on the left edge and one zero mode of flavor $b$ localized on the right edge (when $H_1$ dominates over $H_0$).
To analyze the phase diagram, one thus needs to study the existence of 
a zero mode of flavor $a$ localized on the left edge via the recursion of Eq. \ref{recun} 
\begin{eqnarray}
0 =  K_{m,m}  u_{m}+ K_{m,m+1}  u_{m+1}
\label{recun01}
\end{eqnarray}
This simple recursion between two consecutive coefficients
\begin{eqnarray}
u_{m+1} =  - \frac{K_{m,m}  }{ K_{m,m+1} } u_{m}
\label{recun01bis}
\end{eqnarray}
can be trivially solved for any realization of the random couplings
\begin{eqnarray}
u_{m+1} =  \left( - \frac{K_{m,m}  }{ K_{m,m+1} } \right) ...   \left( - \frac{K_{11}  }{ K_{12} } \right)   u_{1}
= u_1 \prod_{k=1}^m \left( - \frac{K_{k,k}  }{ K_{k,k+1} } \right) 
\label{recun01solu}
\end{eqnarray}
The normalization condition of Eq. \ref{normazero}
\begin{eqnarray}
1= \sum_{n=1}^N u_n^2 = u_1^2 \sum_{n=1}^N \prod_{k=1}^{n-1} \left(  \frac{K^2_{k,k}  }{ K^2_{k,k+1} } \right)
\label{normazero01}
\end{eqnarray}
 corresponds to the well-known structure of Kesten random variables
\cite{Kesten,Der_Pom,Bou,Der_Hil,Cal}
and has been much discussed in relation with the surface magnetization in the ground-state of the one-dimensional 
transverse field Ising chain
\cite{c_microcano,us_watermelon,mblcayley}.

The Lyapunov exponent of this zero mode is then determined by the disorder-average of the logarithms of the couplings
\begin{eqnarray}
\gamma^{\{H_0+H_1\}} \equiv  \lim_{N \to +\infty} \frac{\ln \vert \frac{u_{N+1}}{u_1} \vert}{N} 
= \lim_{N \to +\infty} \frac{ \sum_{k=1}^N \ln \vert \frac{K_{k,k}  }{ K_{k,k+1} } \vert}{N} 
= \overline{ \ln \vert K_{k,k}  \vert - \ln \vert K_{k,k+1}  \vert  }
\label{gamma01}
\end{eqnarray}
and allows to determine if one can construct a localized zero mode near the left edge satisfying the normalization of Eq. \ref{normazero01}.

One obtains the following topological phase diagram :

The phase $n=1$ corresponds to the region of negative Lyapunov exponent $\gamma^{\{H_0+H_1\}} <0$, where the 
normalized zero mode of flavor $a$ is localized near the left edge and typically decays exponentially.

The phase $n=0$ corresponds to the region of positive Lyapunov exponent $\gamma^{\{H_0+H_1\}} >0$, where there is no 
 normalizable zero mode of flavor $a$ localized near the left edge.

The phase transition between the two phases $n=0,1$ corresponds to the vanishing of the Lyapunov exponent 
$\gamma^{\{H_0+H_1\}} =0$,
i.e. to the well-known criterion in the language of the Random Transverse Field Ising Chain \cite{pfeuty,fisher}.
The corresponding Infinite Disorder character of the transition is reviewed in \cite{strong_review,colloquium}.

\subsection{ Example $H=H_0+H_2$  }

\label{subsec02}

For the case
\begin{eqnarray}
H=H_0+H_2= i \sum_{n=1}^N b_n K_{n,n}  a_{n} +  i \sum_{n=1}^{N-2} b_n K_{n,n+2}  a_{n+2}
\label{halpha02}
\end{eqnarray}
(see Eqs \ref{halpha0} and \ref{halpha2} in Appendix \ref{sec_app} for the translation in the spin language),
Eq. \ref{recun} yields one solvable recursion for the odd coefficients
\begin{eqnarray}
u_{2N+1} =  \left( - \frac{K_{2N-1,2N-1}  }{ K_{2N-1,2N+1} } \right) u_{2N-1}
= u_1 \prod_{k=1}^m \left( - \frac{K_{2m-1,2m-1}  }{ K_{2m-1,2m+1} } \right) 
\label{recun02}
\end{eqnarray}
and one solvable recursion for the even coefficients
\begin{eqnarray}
u_{2N+2} =  \left( - \frac{K_{2N,2N}  }{ K_{2N,2N+2} } \right) u_{2N}
= u_2 \prod_{k=1}^m \left( - \frac{K_{2m,2m}  }{ K_{2m,2m+2} } \right) 
\label{recun02solu}
\end{eqnarray}
The corresponding Lyapunov exponents
\begin{eqnarray}
\gamma^{\{H_0+H_2\}}_{odd} =  \lim_{N \to +\infty} \frac{\ln \vert \frac{u_{2N+1}}{u_1} \vert}{2N} 
= \lim_{N \to +\infty} \frac{ \sum_{m=1}^N \ln \vert \frac{K_{2m-1,2m-1}  }{ K_{2m-1,2m+1} }  \vert}{2N} 
=\frac{1}{2}  \overline{ \ln \vert K_{k,k}  \vert - \ln \vert K_{k,k+2}  \vert  }
\label{gamma02odd}
\end{eqnarray}
and
\begin{eqnarray}
\gamma^{\{H_0+H_2\}}_{even} =  \lim_{N \to +\infty} \frac{\ln \vert \frac{u_{2N+2}}{u_2} \vert}{2N} 
= \lim_{N \to +\infty} \frac{ \sum_{m=1}^N \ln \vert \frac{K_{2m,2m}  }{ K_{2m,2m+2} }   \vert}{2N} 
=\frac{1}{2}  \overline{ \ln \vert K_{k,k}  \vert - \ln \vert K_{k,k+2}  \vert  }
\label{gamma02even}
\end{eqnarray}
are thus equal to the value
\begin{eqnarray}
\gamma^{\{H_0+H_2\}} _{odd} = \gamma^{\{H_0+H_2\}}_{even}
=\frac{1}{2}  \overline{ \ln \vert K_{k,k}  \vert - \ln \vert K_{k,k+2}  \vert  } \equiv \gamma^{\{H_0+H_2\}} 
\label{gamma02}
\end{eqnarray}

The phase $n=2$ corresponds to to the region $\gamma^{\{H_0+H_2\}} <0$, where the two zero modes are localized near the left edge and 
display the same typical exponential decay.

The phase $n=0$ corresponds to to the region  $\gamma^{\{H_0+H_2\}} >0$, where there is no localized edge zero mode.

The phase transition between the two phases $n=0,2$ corresponds to 
the simultaneous delocalization transition $\gamma^{\{H_0+H_2\}} =0$ of the two zero modes.

\subsection{ Example $H=H_1+H_2$  }

For the case
\begin{eqnarray}
H=H_1+H_2= i \sum_{n=1}^{N-1} b_n K_{n,n+1}  a_{n+1} +  i \sum_{n=1}^{N-2} b_n K_{n,n+2}  a_{n+2}
\label{halpha12}
\end{eqnarray}
(see Eqs \ref{halpha1} and \ref{halpha2} in Appendix \ref{sec_app} for the translation in the spin language),
Eq. \ref{recun} reads
\begin{eqnarray}
0 =  K_{m,m+1}  u_{m+1}+ K_{m,m+2}  u_{m+2}
\label{recun12}
\end{eqnarray}
So there always exists the trivial zero mode $a_1$ localized on the single site $m=1$, corresponding to the singular Lyapunov exponent
\begin{eqnarray}
\gamma_-^{\{H_0+H_2\}}  = - \infty
\label{gamma02moins}
\end{eqnarray}
while the possible second zero mode has for coefficients
\begin{eqnarray}
u_{N+2} =  \left( - \frac{K_{N,N+1}  }{ K_{N,N+2} } \right) u_{N+1}
= u_2 \prod_{m=1}^N \left( - \frac{K_{m,m+1}  }{ K_{m,m+2} }  \right) 
\label{recun12solu}
\end{eqnarray}
and is thus characterized by the Lyapunov exponent
\begin{eqnarray}
\gamma_+^{\{H_0+H_2\}}  =  \lim_{N \to +\infty} \frac{\ln \vert \frac{u_{N+2}}{u_2} \vert}{N} 
= \lim_{N \to +\infty} \frac{ \sum_{m=1}^N \ln \vert \frac{K_{m,m+1}  }{ K_{m,m+2} }  \vert}{N} 
= \overline{ \ln \vert K_{k,k+1}  \vert - \ln \vert K_{k,k+2}  \vert  }
\label{gamma02plus}
\end{eqnarray}

The phase $n=2$ corresponds to $\gamma^{\{H_1+H_2\}}_+<0$, where the second zero mode is localized near the left edge.

The phase $n=1$ corresponds to $\gamma^{\{H_1+H_2\}}_+ >0$, where one cannot construct a 
second normalized zero mode localized near the left edge.

The phase transition between the two phases $n=1,2$ corresponds to 
the delocalization transition $\gamma^{\{H_1+H_2\}}_+ =0$ of the second zero mode.

\subsection{ Discussion }

As seen on the three examples above, the linear combination $(H_{\alpha_1}+H_{\alpha_2})$ involving only two values of $\alpha$
can be studied via the explicit computation of the possible zero modes for arbitrary couplings,
so that the location of the phase transition between the two possible topological phases are exactly known in the presence of arbitrary disorder.
In the following section, we focus on the case $H=H_0+H_1+H_2$ involving three values of $\alpha$.

\section{ Study of the topological phases of random Hamiltonian  $H=H_0+H_1+H_2$  }

\label{sec_h3}

For the Hamiltonian $H=H_0+H_1+H_2$ of Eq. \ref{h012}
(see Eqs \ref{halpha0}, \ref{halpha1} and \ref{halpha2} in Appendix \ref{sec_app} for the translation in the spin language),
 the recursion equation \ref{recun} for the coefficients of the zero mode
\begin{eqnarray}
0 = K_{m,m}  u_{m}+ K_{m,m+1}  u_{m+1}+ K_{m,m+2}  u_{m+2}
\label{recunh012}
\end{eqnarray}
corresponds to a linear recurrence involving three consecutive terms
\begin{eqnarray}
u_{m+2} = - \frac{K_{m,m+1}  }{ K_{m,m+2} } u_{m+1} - \frac{K_{m,m}  }{ K_{m,m+2} } u_{m} 
\label{recun012}
\end{eqnarray}

\subsection{ Product of random $2 \times 2$ matrices }

It is standard to rewrite the recurrence of Eq. \ref{recun012}
 as
\begin{eqnarray}
 \begin{pmatrix}
u_{m+2}   \\
u_{m+1} 
\end{pmatrix} &&= T_m  \begin{pmatrix}
u_{m+1}   \\
u_{m} 
\end{pmatrix} 
\label{recuntransfert}
\end{eqnarray}
in terms of the $2 \times 2$ transfer matrix
\begin{eqnarray}
T_m =\begin{pmatrix}
- \frac{K_{m,m+1}  }{ K_{m,m+2} }   &  - \frac{K_{m,m}  }{ K_{m,m+2} }   \\
1 &  0
\end{pmatrix}
\label{transfert}
\end{eqnarray}
so that the solution can be obtained from the product of the random transfer matrices
\begin{eqnarray}
 \begin{pmatrix}
u_{N+2}   \\
u_{N+1} 
\end{pmatrix} &&= T_{N} ... T_1  \begin{pmatrix}
u_{2}   \\
u_{1} 
\end{pmatrix} 
\label{prodtransfert}
\end{eqnarray}

Since the product 
\begin{eqnarray}
 {\cal T}_N && \equiv  T_{N} ... T_1  
\label{defprod}
\end{eqnarray}
is a $2 \times 2$ matrix, the product of its two eigenvalues $\tau_{\pm}^{(N)}$ 
can be computed from its determinant as
\begin{eqnarray}
\tau_{+}^{(N)}\tau_{-}^{(N)}=det( {\cal T}_N ) &&  = \prod_{m=1}^N det(T_m) = \prod_{m=1}^N 
\left( \frac{K_{m,m}  }{ K_{m,m+2} } \right)
\label{detprod}
\end{eqnarray}
As a consequence, the two corresponding Lyapunov exponents $\gamma^{\pm}$
\begin{eqnarray}
\gamma_{\pm} \equiv  \lim_{N \to +\infty} \frac{\ln \vert \tau_{\pm}^{(N)} \vert}{N} 
\label{deflyapunov}
\end{eqnarray}
satisfy the simple sum rule \cite{lieu2018}
\begin{eqnarray}
\gamma_{+}+ \gamma_{-} = \lim_{N \to +\infty} \frac{ \ln \vert det( {\cal T}_N ) \vert }{N} 
=  \lim_{N \to +\infty} \frac{ \sum_{m=1}^{N} \ln \vert \frac{K_{m,m}  }{ K_{m,m+2} }  \vert }{N} 
= \overline{ \ln \vert K_{m,m}  \vert - \ln \vert K_{m,m+2}  \vert } 
\label{sumrule}
\end{eqnarray}
with the following consequences \cite{lieu2018} with the ordering $\gamma_- \leq \gamma_+$ :

(i) if $ \overline{ \ln \vert K_{m,m}  \vert - \ln \vert K_{m,m+2}  \vert } <0$, then the smallest Lyapunov exponent has to be negative $\gamma_- <0 $ so there is at least one zero mode,
and the only possibles phases are $n=1,2$ (the phase $n=0$ is excluded), 
while the transition between the two corresponds to the vanishing of the biggest Lyapunov exponent
\begin{eqnarray}
\gamma^{criti(n=1,2)}_+=0
\label{critin12}
\end{eqnarray}

(ii) if $ \overline{ \ln \vert K_{m,m}  \vert - \ln \vert K_{m,m+2}  \vert } >0$, then the biggest Lyapunov exponent has to be positive $\gamma_+>0$, so there is at most one zero mode,
and the only possibles phases are $n=0,1$  (the phase $n=2$ is excluded), while the transition between the two corresponds to 
the vanishing of the smallest Lyapunov exponent
\begin{eqnarray}
\gamma^{criti(n=0,1)}_-=0
\label{critin01}
\end{eqnarray}

(iii) if $ \overline{ \ln \vert K_{m,m}  \vert - \ln \vert K_{m,m+2}  \vert } =0$, then 
either both Lyapunov exponent vanish $\gamma_- = 0 = \gamma_+$ corresponding to the simultaneous delocalization transition of the two zero modes (see the example previously discussed at the end of the subsection \ref{subsec02}),
or the two Lyapunov exponents are opposite $\gamma_- < 0 < \gamma_+=-\gamma_-$ corresponding to the phase $n=1$.

\subsection{ Non-linear recurrence for the Riccati ratio }

Another standard approach \cite{luck,vulpiani,tourigny} 
involves the introduction of the Riccati ratios
\begin{eqnarray}
R_m \equiv \frac{u_{m+1} }{u_m}  
\label{ricc}
\end{eqnarray}
in order to transform the second-order linear recurrence of Eq. \ref{recun012}
into the first-order non-linear recurrence
\begin{eqnarray}
R_{m+1} 
=  - \frac{K_{m,m+1}  }{ K_{m,m+2} } - \frac{K_{m,m}  }{ K_{m,m+2} } \frac{1}{R_m}
\label{recricc}
\end{eqnarray}

In terms of these Riccati ratios,
the biggest Lyapunov exponent reads
\begin{eqnarray}
\gamma_+ =  \lim_{N \to +\infty} \frac{\ln \vert \frac{u_{N+1}}{u_1} \vert}{N} 
= \lim_{N \to +\infty} \frac{ \sum_{m=1}^N \ln \vert R_m \vert}{N} 
= \int_{-\infty}^{+\infty} dR  \  {\cal P}_{st}(R) \ln \vert R \vert
\label{gammaplusstatio}
\end{eqnarray}
where $ {\cal P}_{st}(R) $ denotes the attractive stationary distribution for the recurrence of Eq. \ref{recricc}.

This formulation in terms of the Riccati ratios is also useful to characterize the finite-size fluctuations
via the Central-Limit-Theorem
\begin{eqnarray}
\ln \left\vert \frac{u_{N+1}}{u_1} \right\vert
= \sum_{m=1}^N \ln \vert R_m \vert 
\opsimeq_{N \to +\infty}  \gamma_+ N  + \sqrt{N}  u
\label{cltlyapunov}
\end{eqnarray}
where $u$ is a Gaussian variable of zero mean and of variance given by the variance of $(\ln \vert R \vert)$
computed with the stationary distribution $ {\cal P}_{st}(R)$.
As a consequence, the phase transition corresponds to an Infinite Disorder Fixed point,
where the typical correlation exponent $\nu_{typ}=1$ characterizes the vanishing of $\gamma_+$ as a function of the control parameter, while the average correlation exponent $\nu_{av}=2$ characterizes the sample-to-sample fluctuations,
as in the much studied Random Transverse Field Ising Chain (see the review \cite{strong_review}).
Hence Strong Disorder Renormalization has been used to analyze the critical points and the Griffiths effects 
for this type of models in the langage of dirty superconductors \cite{motrunich}.

\subsection{ Reminder on the numerical results for the case of random couplings $K_{mm}$ \cite{lieu2018} }

In the presence of arbitrary disorder, recurrences like Eq. \ref{recricc} are not exactly soluble
and are usually studied numerically, for instance the results for the case with random $K_{m,m}$ and non-random $K_{m,m+1}$ and $K_{m,m+2}$ can be found in Ref. \cite{lieu2018}, with the following conclusions for the phase diagram :

(a) The direct phase transition between the phases $n=0$ and $n=2$ that requires the simultaneous delocalization of the two zero modes
 $\gamma_- = 0 = \gamma_+$ is possible only in the absence of disorder,
while the presence of disorder induces a splitting between the two Lyapunov exponents
and thus introduces an intermediate phase $n=1$ even for arbitrary weak disorder.
More generally, the Lyapunov spectrum is expected to be non-degenerate in random systems
so that the phase transitions only change the topological index by one \cite{motrunich}.

(b) The presence of disorder may favor the existence of Majorana Zero Mode,
i.e. in the phase diagram, a point corresponding to the phase $n=0$ for the pure model may belong to the phase $n=1$ in the presence of sufficient disorder, or similarly a point corresponding to the phase $n=1$ for the pure model may belong to the phase $n=2$ in the presence of sufficient disorder.

(c) As the disorder becomes very strong for the couplings $K_{m,m}$, one expects transition $n=2 \to n=1 \to n=0$ towards the trivial phase $n=0$.

In the following, it is thus interesting to analyze the same questions for the phase diagram
corresponding to another type of disorder, namely in the coupling $K_{m,m+1}$.

\section{ Explicit solution for $H=H_0+H_1+H_2$ with Cauchy disorder in $K_{m,m+1}$ }

\label{sec_cauchy}

Among the exactly soluble cases of first-order non-linear recurrence for Riccati ratios \cite{luck,vulpiani,tourigny},
the simplest explicit case is the Lloyd model \cite{lloyd,thouless}, as already used in the context of random Majorana models
 in \cite{degottardi2012}.
For our present model, this soluble case for the Riccati recurrence  
corresponds the case where the couplings $K_{m,m} $ and $ K_{m,m+2} $ are non-random
\begin{eqnarray}
K_{m,m} =k_0
\nonumber \\
K_{m,m+2} =k_2
\label{k0k2fixed}
\end{eqnarray}
while the couplings $K_{m,m+1}$ are distributed with the Cauchy distribution $C_{k_1,W}$ of average $k_1$ and broadness $W$
\begin{eqnarray}
C_{k_1,W} (K_{m,m+1}) = \frac{ 1}{\pi} \ \frac{W}{ (K_{m,m+1}-k_1)^2+W^2}
\label{cauchy}
\end{eqnarray}

\subsection{ Recurrence for the parameters of the Cauchy distribution }

Then the recurrence of Eq. \ref{recricc} simplifies into
\begin{eqnarray}
R_{m+1} 
=  - \frac{ K_{m,m+1} }{ k_2 } - \frac{ k_0  }{ k_2} \frac{1}{R_m}
\label{recricccauchy}
\end{eqnarray}
Since the Cauchy distribution is stable with respect to both addition and inversion,
the Riccati ratios $R_m$ are then also distributed with the Cauchy distribution $C_{x_m,y_m}$ with some average $x_m$
 and some broadness $y_m>0$
\begin{eqnarray}
C_{x_m,y_m} (R_m) = \frac{ 1}{\pi} \ \frac{y_m}{ (R_m-x_m)^2+y_m^2}
\label{cauchym}
\end{eqnarray}
with the Fourier transform
\begin{eqnarray}
\int_{-\infty}^{+\infty} dR_{m} e^{i q R_m} C_{x_m,y_m} (R_m) =e^{i q x_m - \vert q \vert y_m} 
\label{fouriercauchy}
\end{eqnarray}

The recurrence of Eq. \ref{recricccauchy} yields the following recurrence for the Cauchy distributions
\begin{eqnarray}
C_{x_{m+1},y_{m+1}} (R_{m+1}) = \int dK_{m,m+1} C_{k_1,W} (K_{m,m+1}) \int dR_{m} C_{x_{m},y_{m}} (R_{m}) 
\delta\left( R_{m+1}  +\frac{ K_{m,m+1} }{ k_2 } + \frac{ k_0  }{ k_2} \frac{1}{R_m} \right)
\label{cauchyrec}
\end{eqnarray}
or equivalently in Fourier transform
\begin{eqnarray}
e^{i q x_{m+1} - \vert q \vert y_{m+1}} 
&& = \int dR_{m+1} e^{i \omega R_{m+1}} C_{x_{m+1},y_{m+1}} (R_{m+1})
\nonumber \\
&& = 
\int dK_{m,m+1} C_{k_1,W} (K_{m,m+1})  e^{- i  \frac{ q }{ k_2} K_{m,m+1}} 
\int dR_{m} C_{x_{m},y_{m}} (R_{m})   e^{- i q \frac{ k_0  }{ k_2} \frac{1}{R_m}} 
\nonumber \\
&& =e^{- i  \frac{q  }{ k_2 } k_1 - \vert  \frac{q  }{ k_2 } \vert W} 
 e^{-i q \frac{ k_0  }{ k_2} \frac{x_{m} } {x_{m}^2+y_{m}^2 } - \vert  \frac{q k_0  }{ k_2 } \vert 
 \frac{y_{m} } {x_{m}^2+y_{m}^2 }}
\label{recfouriercauchy}
\end{eqnarray}
The recurrence for the average $x_m$ and the broadness $y_m$ thus reads
\begin{eqnarray}
x_{m+1} && =-  \frac{ k_1  }{ k_2 } -  \frac{ k_0  }{ k_2 } \frac{x_{m} } {x_{m}^2+y_{m}^2 }
\nonumber \\
y_{m+1} && = \frac{ W  }{ \vert k_2 \vert } + \left\vert  \frac{ k_0  }{ k_2 } \right\vert 
 \frac{y_{m} } {x_{m}^2+y_{m}^2 }
\label{cauchyrecxy}
\end{eqnarray}
and the corresponding attractive fixed point $(x_f,y_f)$ satisfies
\begin{eqnarray}
x_f && =-  \frac{ k_1  }{ k_2 } -  \frac{ k_0  }{ k_2 } \frac{x_f } {x_f^2+y_f^2 }
\nonumber \\
y_f && = \frac{ W  }{ \vert k_2 \vert } + \left\vert  \frac{ k_0  }{ k_2 } \right\vert  \frac{y_f } {x_f^2+y_f^2 }
\label{cauchyrecxyf}
\end{eqnarray}

From the stationary distribution ${\cal P}_{st}(R) =C_{x_f,y_f}(R)$ of the Riccati ratio $R$,
Eq. \ref{gammaplusstatio} yields
the biggest Lyapunov exponent 
\begin{eqnarray}
\gamma_{+} = \int_{-\infty}^{+\infty} dR  {\cal P}_{st}(R) \ln \vert R \vert
= \int dR  \frac{ 1}{\pi} \ \frac{y_f}{ (R-x_f)^2+y_f^2}   \ln \vert R \vert
= \frac{ \ln (x_f^2+y_f^2)  }{2 }
\label{gammapres}
\end{eqnarray}
while the sum rule of Eq. \ref{sumrule} yields the other Lyapunov exponent
\begin{eqnarray}
 \gamma_{-} = \ln \left \vert \frac{ k_0}{k_2} \right \vert - \gamma_{+}
\label{sumrulem}
\end{eqnarray}

The solution of the fixed-point system of Eq. \ref{cauchyrecxyf} depends on the sign of the ratio $ \frac{ k_0  }{ k_2 } $,
i.e. whether the two couplings $k_0$ and $k_2$ have the same sign or not.
These two cases are thus analyzed in the two following subsections respectively.

\subsection{  Lyapunov exponents for  the case $ \frac{ k_0  }{ k_2}  = - \left\vert  \frac{ k_0  }{k_2 } \right\vert   <0$ }

When the two couplings $k_0$ and $k_2$ have opposite signs, Eq \ref{cauchyrecxyf} can be rewritten as
\begin{eqnarray}
x_f  && =- \frac{ \frac{ k_1  }{ k_2 } }{1 - \left\vert  \frac{ k_0  }{k_2 } \right\vert   \frac{1} {x_f^2+y_f^2 }}
\nonumber \\
y_f && =\frac{ \frac{ W  }{ \vert k_2 \vert }}{1- \left\vert  \frac{ k_0  }{k_2 } \right\vert   \frac{1} {x_f^2+y_f^2 }}
\label{cauchyrecxyfneg}
\end{eqnarray}
It is then convenient to introduce the polar coordinates
\begin{eqnarray}
x_f  && =r_f \cos \theta_f
\nonumber \\
y_f && =r_f \sin \theta_f
\label{polar}
\end{eqnarray}
Since the broadness is positive $y_f>0$, 
the angle belongs to $\theta \in [0,\pi[$ and is determined by its tangent
\begin{eqnarray}
\tan \theta_f = \frac{y_f}{x_f}   && =-  \frac{W  k_2  }{ k_1 \vert k_2 \vert} 
\label{tangent}
\end{eqnarray}
while the modulus $r_f$ satisfies the bound $\left(1- \left\vert  \frac{ k_0  }{k_2 } \right\vert   \frac{1} {r_f^2 } \right) >0$ 
and thus the equation
\begin{eqnarray}
r_f = \sqrt{x_f^2+y_f^2} =  \frac{ \frac{ \sqrt{k_1^2+W^2}  }{ \vert k_2 \vert }}{1- \left\vert  \frac{ k_0  }{k_2 } \right\vert   \frac{1} {r_f^2 }}
\label{eqmodulus}
\end{eqnarray}
The rewriting as
a second order equation
\begin{eqnarray}
r_f^2 - \frac{\sqrt{k_1^2+W^2}}{\vert k_2 \vert} r_f  - \left\vert  \frac{ k_0  }{k_2 } \right\vert  =0 
\label{2emedegrerf}
\end{eqnarray}
yields that the only positive solution reads
\begin{eqnarray}
r_f 
% = \frac{  \frac{\sqrt{k_1^2+W^2}}{\vert k_2 \vert} + \sqrt{ \frac{k_1^2+W^2}{ k_2^2 } +4 \left\vert  \frac{ k_0  }{k_2 } \right\vert }   }{2 }
=  \frac{  \sqrt{k_1^2+W^2} + \sqrt{ k_1^2+W^2 +4 \vert  k_0  k_2 \vert }   }{2 \vert k_2 \vert}
\label{solurff}
\end{eqnarray}

The biggest Lyapunov exponent of Eq \ref{gammapres} becomes
\begin{eqnarray}
\gamma_{+} = \frac{ \ln (x_f^2+y_f^2)  }{2 } = \ln r_f 
= \ln \left(  \frac{  \sqrt{k_1^2+W^2} + \sqrt{ k_1^2+W^2 +4 \vert  k_0  k_2 \vert }   }{2 \vert k_2 \vert} \right)
\label{gammapresm}
\end{eqnarray}
while the other Lyapunov exponent $\gamma_{-} \leq \gamma_{+} $ can be obtained from the sum of Eq. \ref{sumrule}
\begin{eqnarray}
\gamma_{-} =   \ln \vert k_0  \vert - \ln \vert k_2  \vert  -\gamma_{+}
= \ln  \left(  \frac{2 \vert k_0 \vert}{  \sqrt{k_1^2+W^2} + \sqrt{ k_1^2+W^2 +4 \vert  k_0  k_2 \vert }   } \right)
\label{gammamresm}
\end{eqnarray}

\subsubsection{ Phase $n=0$ }

The phase $n=0$ corresponds to two positive Lyapunov exponents $0< \gamma^-  (\leq \gamma^+ )$, i.e. to the region
\begin{eqnarray}
 \sqrt{k_1^2+W^2} + \sqrt{ k_1^2+W^2 +4 \vert  k_0  k_2 \vert }    < 2 \vert k_0 \vert 
\label{phasen0m}
\end{eqnarray}
which can be simplified into
\begin{eqnarray}
 \sqrt{k_1^2+W^2} + \vert k_2 \vert     <  \vert k_0 \vert  
\label{phasen0mbis}
\end{eqnarray}

\subsubsection{ Phase $n=2$ } 

The phase $n=2$ corresponds to two negative Lyapunov exponents $(\gamma^- \leq ) \gamma^+ <0$, i.e. to the region
\begin{eqnarray}
  \sqrt{k_1^2+W^2} + \sqrt{ k_1^2+W^2 +4 \vert  k_0  k_2 \vert }    < 2 \vert k_2 \vert 
\label{phasen2m}
\end{eqnarray}
which can be simplified into
\begin{eqnarray}
 \sqrt{k_1^2+W^2} + \vert k_0 \vert     <  \vert k_2 \vert  
\label{phasen2mbis}
\end{eqnarray}

\subsubsection{ Phase $n=1$ } 

Finally, the phase $n=1$ corresponds to the case $\gamma^- <0< \gamma^+ $, i.e. to the region
\begin{eqnarray}
( \vert k_2 \vert  - \vert k_0 \vert  )^2 < k_1^2+W^2
\label{phasen1m}
\end{eqnarray}

\subsubsection{ Conclusion : Locations of the phase transitions between the three phases in the region $ \frac{ k_0  }{k_2 }  = -\left\vert  \frac{ k_0  }{k_2} \right\vert   <0$} 

The critical line between the phases $n=0$ and $n=1$ corresponds to
\begin{eqnarray}
 \sqrt{k_1^2+W^2} + \vert k_2 \vert     =  \vert k_0 \vert  
\label{criti01m}
\end{eqnarray}

The critical line between the phases $n=1$ and $n=2$ corresponds to
\begin{eqnarray}
 \sqrt{k_1^2+W^2} + \vert k_0 \vert     =  \vert k_2 \vert  
\label{criti12m}
\end{eqnarray}

A direct transition between the phases $n=0$ and $n=2$ requires the condition
\begin{eqnarray}
 \sqrt{k_1^2+W^2} =  \vert k_0 \vert  -\vert k_2 \vert   = \vert k_2 \vert  -\vert k_0 \vert
\label{criti02m}
\end{eqnarray}
which can be fulfilled only for the case  $ \vert k_0 \vert  = \vert k_2 \vert $ and $(k_1=0,W=0)$ where the couplings $K_{m,m+1}$ all vanish (this case has been previously discussed in the subsection \ref{subsec02}).

\subsection{ Lyapunov exponents for the case $ \frac{ k_0  }{k_2 }  = +\left\vert  \frac{ k_0  }{k_2} \right\vert   >0$ }

When the two couplings $k_0$ and $k_2$ have the same sign, 
it is convenient to recast the system of Eq \ref{cauchyrecxyf} 
\begin{eqnarray}
x_f && =-  \frac{ k_1  }{ k_2 } -   \frac{ k_0  }{k_2}  \frac{x_f } {x_f^2+y_f^2 }
\nonumber \\
y_f && = \frac{ W  }{ \vert k_2 \vert } +   \frac{ k_0  }{ k_2 }   \frac{y_f } {x_f^2+y_f^2 }
\label{cauchyrecxyfp}
\end{eqnarray}
into the following single equation for the complex variable $z_f=x_f+i y_f$
\begin{eqnarray}
z_f  && =   - \omega-    \frac{ k_0  }{k_2} \frac{1} {z_f }
\label{cauchyreczfp}
\end{eqnarray}
where we have introduced the notation
\begin{eqnarray}
\omega && \equiv   \frac{ k_1  }{ k_2 }  -i \frac{ W  }{ \vert k_2 \vert } 
\label{omega}
\end{eqnarray}
Rewriting Eq \ref{cauchyrecxyfp}
as a second degree equation
\begin{eqnarray}
z_f^2 +\omega z_f +   \frac{ k_0  }{k_2}  =0 
\label{zfsecond}
\end{eqnarray}
one obtains the two solutions
\begin{eqnarray}
z_{f\pm} = \frac{ - \omega \pm \sqrt{\Delta} }{2}
\label{soluzfpm}
\end{eqnarray}
in terms of the discriminant
\begin{eqnarray}
\Delta= \omega^2 -4  \frac{ k_0  }{k_2}
%  =  \frac{ k_1^2 - W^2 -4 \vert k_0 k_2 \vert }{ k_2^2 }  -i 2 \frac{ k_1  }{ k_2 } \frac{ W  }{ \vert k_2 \vert } 
\label{delta}
\end{eqnarray}

Since we are interested into the squares of the modulus of the solutions, it is actually convenient to use 
that their product reads
\begin{eqnarray}
\vert z_{f+}\vert^2 \vert z_{f-} \vert^2 && =  \frac{ k_0^2  }{k_2^2} 
\label{prodsum}
\end{eqnarray}
while their sum is given by
\begin{eqnarray}
S \equiv \vert z_{f+}\vert^2 + \vert z_{f-} \vert^2 && 
% =\left\vert \frac{ - \omega + \sqrt{\Delta} }{2} \right\vert^2 +\left\vert \frac{ - \omega - \sqrt{\Delta} }{2} \right\vert^2
= \frac{\vert \omega^2 \vert + \vert \Delta \vert}{2}
= \frac{\vert \omega^2 \vert + \left\vert 
 \omega^2 -4   \frac{ k_0  }{k_2} 
 \right\vert}{2}
% = \frac{ k_1^2+W^2 +  \sqrt{ (k_1^2 - W^2 -4 \vert k_0 k_2 \vert )^2 + 4 k_1^2 W^2 } }{2 k_2^2} 
\label{sommesquare}
\end{eqnarray}
So $r_f^2=\vert z_f \vert^2$ satisfies the second order equation
\begin{eqnarray}
r_f^4-  S r_f^2 + \frac{ k_0^2  }{k_2^2}  =0
\label{eqr2}
\end{eqnarray}
of discriminant
\begin{eqnarray}
D = S^2 -4 \frac{ k_0^2  }{k_2^2}  
&& = \frac{ \left( \vert \omega^2 \vert + \left\vert   \omega^2 -4  \frac{ k_0  }{k_2}   \right\vert \right)^2
 - 16 \frac{ k_0^2  }{k_2^2} }{4}
\nonumber \\
&& = 
\frac{ \left( \vert \omega^2 \vert + \left\vert   \omega^2 -4  \frac{ k_0  }{k_2}  \right\vert 
+4   \frac{ k_0  }{k_2} \right)
\left( \vert \omega^2 \vert + \left\vert   \omega^2 -4   \frac{ k_0  }{k_2}   \right\vert 
-4  \frac{ k_0  }{k_2} \right)
 }{4}
\label{discri}
\end{eqnarray}
which is positive (as it should to have real roots) as a consequence of the triangular inequality.
From the two solutions 
\begin{eqnarray}
r_{f\pm}^2 && =\frac{ S \pm \sqrt{ D }  }{2}
\label{solur2}
\end{eqnarray}
one obtains that the biggest Lyapunov exponent reads
\begin{eqnarray}
\gamma_{+} =  \frac{ \ln (r_{f+}^2)  }{2 }=  \frac{ \ln \left( \frac{ S + \sqrt{ D }  }{2} \right) }{2 }
\label{gammapressolu}
\end{eqnarray}
while the smallest reads (Eq. \ref{sumrule})
\begin{eqnarray}
\gamma_{-} =   \ln \vert k_0  \vert - \ln \vert k_2  \vert  -\gamma_{+}
= \frac{ \ln \frac{
\frac{ k_0^2  }{k_2^2} 
}{(r_{f+}^2)}   }{2 }
=  \frac{ \ln (r_{f-}^2)  }{2 }=  \frac{ \ln \left( \frac{ S - \sqrt{D}  }{2} \right) }{2 }
\label{gammamresp}
\end{eqnarray}

In terms of the initial parameters, Eq \ref{sommesquare} reads
\begin{eqnarray}
S  && 
 = \frac{ k_1^2+W^2  +  \sqrt{ [ (k_1+2 \sqrt{k_0 k_2} )^2+W^2 ] [ (k_1-2 \sqrt{k_0k_2})^2+W^2 ] }  }{2 k_2^2} 
\nonumber \\
&&  = \frac{ k_1^2+W^2  +  \sqrt{ ( k_1^2-4 k_0 k_2 )^2+2 W^2 (k_1^2+4 k_0 k_2) +W^4 }   }{2 k_2^2} 
\label{sommeexplicit}
\end{eqnarray}

\subsubsection{ Phase $n=2$ }

The phase $n=2$ corresponds to two negative Lyapunov exponents $(\gamma^- \leq ) \gamma^+ <0$, i.e. to the region
\begin{eqnarray}
 S < 1+ \frac{k_0^2}{k_2^2} <2
\label{sphasen2p}
\end{eqnarray}
leading finally to the two conditions
\begin{eqnarray}
&& \vert k_0 \vert < \vert k_2 \vert 
%  \nonumber \\ && k_1^2+W^2 < 2 (k_0^2+k_2^2)
\nonumber \\
&& k_1^2 (k_2-k_0)^2 + W^2 (k_2+k_0)^2 < (k_2-k_0)^2 (k_2+k_0)^2
\label{phasen2p}
\end{eqnarray}

For the non-random case $W=0$, these two conditions reduce to
\begin{eqnarray}
&& \vert k_0 \vert < \vert k_2 \vert 
 \nonumber \\
&& \vert k_1 \vert <  \vert k_2+k_0 \vert
\label{phasen2p2condw0}
\end{eqnarray}

\subsubsection{ Phase $n=1$ }

The phase $n=1$ corresponds to the case $\gamma^- <0< \gamma^+ $, i.e. to the region
\begin{eqnarray}
S > 1+ \frac{k_0^2}{k_2^2} 
\label{phasen1p}
\end{eqnarray}
that finally leads to the condition
\begin{eqnarray}
&& (k_2-k_0)^2 (k_2+k_0)^2 < k_1^2 (k_2-k_0)^2 + W^2 (k_2+k_0)^2 
\label{phasen1p2cond}
\end{eqnarray}

For the non-random case $W=0$, 
Eq \ref{phasen1p2cond} reduces to
\begin{eqnarray}
  \vert k_2+k_0 \vert < \vert k_1 \vert 
\label{phasen1p2condw0}
\end{eqnarray}

\subsubsection{ Phase $n=0$ } 

The phase $n=0$ corresponds to two positive Lyapunov exponents $0< \gamma^-  (\leq \gamma^+ )$, i.e. to the region
\begin{eqnarray}
2 < S < 1+ \frac{k_0^2}{k_2^2} 
\label{phasen0p}
\end{eqnarray}
leading finally to the two conditions
\begin{eqnarray}
&& \vert k_2 \vert < \vert k_0 \vert 
%  \nonumber \\ && k_1^2+W^2 < 2 (k_0^2+k_2^2)
\nonumber \\
&& k_1^2 (k_2-k_0)^2 + W^2 (k_2+k_0)^2 < (k_2-k_0)^2 (k_2+k_0)^2
\label{phasen0p2cond}
\end{eqnarray}

For the non-random case $W=0$, these two conditions reduce to
\begin{eqnarray}
&& \vert k_2 \vert < \vert k_0 \vert 
 \nonumber \\
&& \vert k_1 \vert <  \vert k_2+k_0 \vert
\label{phasen0p2condw0}
\end{eqnarray}

\subsubsection{ Conclusion : Location of the phase transition between the three phases in the region $ \frac{ k_0  }{k_2 }  = +\left\vert  \frac{ k_0  }{k_2} \right\vert   >0$} 

The critical line between the phases $n=1$ and $n=2$ corresponds to
to the two conditions
\begin{eqnarray}
&& \vert k_0 \vert < \vert k_2 \vert 
\nonumber \\
&& k_1^2 (k_2-k_0)^2 + W^2 (k_2+k_0)^2 = (k_2-k_0)^2 (k_2+k_0)^2
\label{cririn1n2p}
\end{eqnarray}
For the non-random case $W=0$, this reduces to
\begin{eqnarray}
&& \vert k_0 \vert < \vert k_2 \vert 
 \nonumber \\
&& \vert k_1 \vert =  \vert k_2+k_0 \vert
\label{cririn1n2pw0}
\end{eqnarray}

The critical line between the phases $n=0$ and $n=1$ corresponds to
to the two conditions
\begin{eqnarray}
&& \vert k_2 \vert < \vert k_0 \vert 
\nonumber \\
&& k_1^2 (k_2-k_0)^2 + W^2 (k_2+k_0)^2 = (k_2-k_0)^2 (k_2+k_0)^2
\label{critin0n1p}
\end{eqnarray}
For the non-random case $W=0$, this reduces to
\begin{eqnarray}
&& \vert k_2 \vert < \vert k_0 \vert 
\nonumber \\
&& \vert k_1 \vert =  \vert k_2+k_0 \vert
\label{critin0n1pw0}
\end{eqnarray}

A direct transition between the phases $n=0$ and $n=2$ requires the condition
\begin{eqnarray}
S = 1+ \frac{k_0^2}{k_2^2} = 2
\label{criti02p}
\end{eqnarray}
Since in the present section the two couplings $k_0$ and $k_2$ have the same sign, 
one obtains the conditions
\begin{eqnarray}
&&  k_0 = k_2
\nonumber \\
&&  W= 0
 \nonumber \\
&& k_1^2 < 2 (k_0^2+k_2^2) = 4 k_0^2
\label{critin0n2p}
\end{eqnarray}
i.e. it is only possible in the absence of disorder $W=0$.

\subsection{ Figures of the phase diagram in the plane $(\frac{k_0}{k_2},\frac{k_1}{k_2})$ for various disorder strengths $\frac{W}{\vert k_2 \vert }$ }

It is now interesting to draw the phase diagram in the plane of the reduced variables
\begin{eqnarray}
x && = \frac{k_0}{k_2}
\nonumber \\
y &&  = \frac{k_1}{k_2}
\label{planexy}
\end{eqnarray}
in order to see how it evolves as a function of the reduced disorder strength
\begin{eqnarray}
w  = \frac{W}{\vert k_2 \vert }
\label{wred}
\end{eqnarray}

\begin{figure}[ht]
  \begin{center}
    \includegraphics[width=17cm]{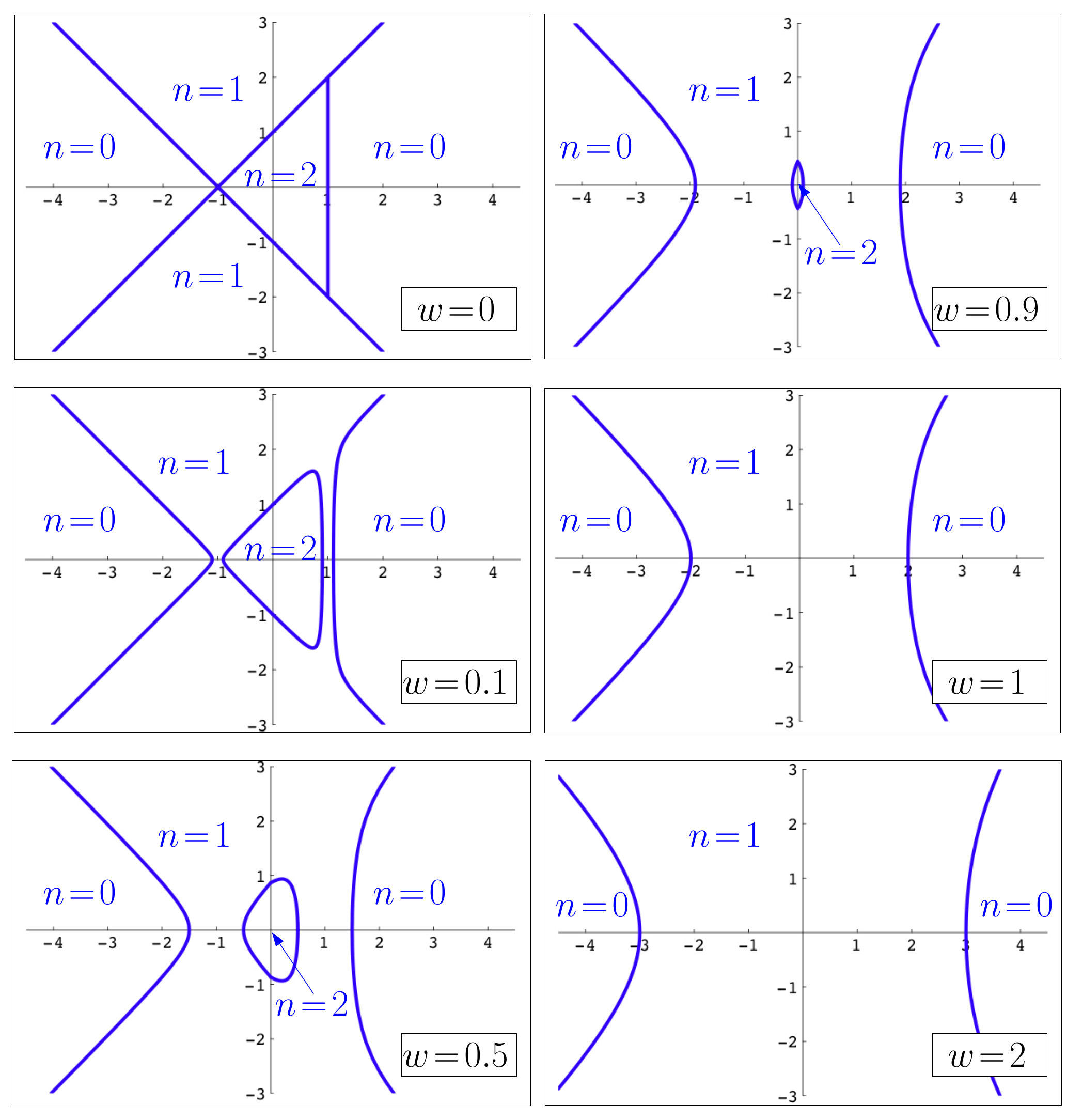}
  \end{center}
%  \vskip -4cm
  \caption{Phase diagram of the three topological phases $n=0,1,2$
in the plane $(x=\frac{k_0}{k_2},y=\frac{k_1}{k_2})$ for various disorder strengths $w=\frac{W}{\vert k_2 \vert }$. 
Note that the direct transition between $n=0$ and $n=2$ occurs only in the non-random case $w=0$ and disappears
for any arbitrary disorder via an intermediate phase $n=1$, as shown with the examples $w=0.1$, $w=0.5$ and $w=0.9$.
 For sufficiently strong disorder $w \geq 1$, the phase $n=2$ cannot exist anymore, as shown with the examples $w=1$ and $w=2$.}
  \label{figure}
\end{figure}

For the non-random case $w=0$, the phase diagram has been already discussed in previous works
\cite{pollmann_unified,pollmann_topology,niu2012,ardonne2018,lieu2018}
and is drawn as the first picture on Figure \ref{figure} as a comparison with the random cases $w>0$.

\subsubsection{ Locations of the phase transitions between the three phases in the region $ x= \frac{ k_0  }{k_2 }  <0$} 

In the half-plane $x<0$, 
the critical line between the phases $n=0$ and $n=1$ corresponds to the full hyperbola branch (Eq \ref{criti01m}) 
\begin{eqnarray}
&& y = \pm \sqrt{ (x+1)^2 -w^2 }
\nonumber \\
&& x \leq -1 - w
\label{criti01mhyp}
\end{eqnarray}
The critical line between the phases $n=1$ and $n=2$ corresponds to the truncated other branch of the same hyperbola
\begin{eqnarray}
&& y = \pm \sqrt{ (x+1)^2 -w^2 }
\nonumber \\
&&  -1 + w \leq x \leq 0
\label{criti12mhyp}
\end{eqnarray}
that exists only for sufficiently small reduced disorder $w<1$ (see Fig \ref{figure}).

In the non-random case $w=0$, this hyperbola degenerate into the two straight lines $y = \pm (x+1)$ (see Fig \ref{figure}) :
the direct transition between the phases $n=0$ and $n=2$ is then reduced to their intersection point $(x=-1,y=0)$.

\subsubsection{ Location of the phase transitions between the three phases in the region $ x=\frac{ k_0  }{k_2 }    >0$} 

In the half-plane $x>0$, the critical line between the phases $n=1$ and $n=2$ corresponds to the truncated branch
\begin{eqnarray}
&& y = \pm (1+x) \sqrt{ 1 -\frac{w^2}{(1-x)^2} }
\nonumber \\
&& 0 \leq x \leq 1-w
\label{critin1n2phyp}
\end{eqnarray}
that exists only for sufficiently small reduced disorder $w<1$ (see Fig \ref{figure}).
The critical line between the phases $n=0$ and $n=1$ corresponds to
to the full other branch of the same curve
\begin{eqnarray}
&& y = \pm (1+x) \sqrt{ 1 -\frac{w^2}{(1-x)^2} }
\nonumber \\
&& 1+w \leq x 
\label{critin0n1phyp}
\end{eqnarray}

For the non-random case $w=0$, the curve above degenerates into the straight lines $ y = \pm (1+x) $
and the direct transition between the phases $n=0$ and $n=2$ becomes possible along the vertical segment (see Fig \ref{figure})
\begin{eqnarray}
&&   -2 \leq y \leq +2
\nonumber \\
&&  x=1
\label{critin0n2pw0}
\end{eqnarray}

\section{  Conclusions }

\label{sec_conclusion}

In this paper, we have considered the topological phase transitions in random Kitaev $\alpha$-chains.
We have first recalled how the edge Majorana Zero Modes could be computed for any realization of disorder for Hamiltonians involving only two values of $\alpha$. We have then focused on the random Hamiltonian $(H_{0} +H_1+H_2 )$
containing three values of $\alpha$, where the localization properties of the edge Majorana Zero Modes
can be analyzed via the product of $2 \times 2$ random matrices and via the Riccati non-linear recurrence. 
For the special case of Cauchy disorder in the couplings $K_{m,m+1}$, we have computed
explicitly the two Lyapunov exponents in order to analyze how the phase diagram of the three topological phases $n=0,1,2$
evolves as a function of the disorder strength.
In particular, we have obtained that the direct phase transition between the phases $n=0$ and $n=2$ 
becomes impossible in the presence of disorder that always induces an intermediate phase $n=1$, 
as found previously via numerics for other distributions of disorder \cite{lieu2018},
and in agreement with the more general expectation
that topological phase transitions in random systems only change the topological index by one 
as a consequence of the non-degeneracy of the Lyapunov spectrum \cite{motrunich}.
We have also obtained that the phase $n=2$ completely disappears for strong enough disorder 
($w \geq 1$ in Figure \ref{figure}).

\appendix

\section{ Dictionary between Majorana fermions and quantum spin chains } 

\label{sec_app}

For a chain of $N$ quantum spins described by Pauli matrices, 
the $(2N)$ string operators 
\begin{eqnarray}
a_j =\gamma_{2j-1} && \equiv \left( \prod_{k=1}^{j-1} \sigma_k^z \right) \sigma_j^x
\nonumber \\
b_j =\gamma_{2j}  && \equiv  \left( \prod_{k=1}^{j-1} \sigma_k^z \right)  \sigma_j^y 
\label{sigmaxy}
\end{eqnarray}
 satisfy  the Majorana anticommutation relations 
\begin{eqnarray}
 \gamma_k \gamma_l + \gamma_l \gamma_k && = 2 \delta_{kl}
\label{anticomm}
\end{eqnarray}

The first examples of Kitaev $\alpha$-chains of Eq. \ref{halpha}
reads in the spin language
\begin{eqnarray}
 H_{(\alpha=0)}= i \sum_m b_m K_{m,m}  a_{m} = \sum_m K_{m,m} \sigma_m^z
\label{halpha0}
\end{eqnarray}
\begin{eqnarray}
 H_{(\alpha=1)}= i \sum_m b_m K_{m,m+1}  a_{m+1} = - \sum_m K_{m,m+1} \sigma_m^x \sigma_{m+1}^x 
\label{halpha1}
\end{eqnarray}
and
\begin{eqnarray}
 H_{(\alpha=2)}= i \sum_m b_m K_{m,m+2}  a_{m+2} = - \sum_m K_{m,m+2} \sigma_m^x \sigma_{m+1}^z \sigma_{m+2}^x 
\label{halpha2}
\end{eqnarray}

\end{document}